\documentclass{sf2a-conf2016}
\usepackage{graphicx}
\usepackage{hyperref}
\usepackage[applemac]{inputenc}
\usepackage[]{natbib}  
\usepackage{epstopdf}

\def\BibTeX{{\rm B\kern-.05em{\sc i\kern-.025em b}\kern-.08em
    T\kern-.1667em\lower.7ex\hbox{E}\kern-.125emX}}
\bibpunct{(}{)}{;}{a}{}{,}  

\begin{document}

\TitreGlobal{SF2A 2016}


\title{Non-thermal emission and dynamical state of massive galaxy clusters from $CLASH$ sample}

\runningtitle{NT and dynamical state of CLASH clusters}

\author{M. Pandey-Pommier}\address{Univ Lyon1, ENS Lyon, CNRS, Centre de Recherche Astrophysique de Lyon UMR5574, 9 av Charles Andr\'e, F- 69230, Saint-Genis-Laval, France}

\author{J. Richard$^1$}

\author{F. Combes}\address{LERMA, 61 Av. de l'Observatoire, F-75 014, Paris, France}

\author{A. Edge}\address{Department of Physics, University of Durham, South Road, Durham DH1 3LE}

\author{B. Guiderdoni$^1$}

\author{D. Narasimha}\address{Theoretical Astrophysics Group, Tata Institute 
of Fundamental Research, Homi Bhabha Road, 400 005 Bombay, India}

\author{J. Bagchi}\address{Inter University Centre for Astronomy and Astrophysics, Pune University Campus, Postbag 4, Pune 411007, 
India}

\author{J. Jacob}\address{Newmann College, Thodupuzha, Kerala, 685584, India}




\setcounter{page}{237}


\maketitle


\begin{abstract}
Massive galaxy clusters are the most violent large scale structures undergoing merger events in the Universe. Based upon their morphological properties in X-rays, 
they are classified as un-relaxed and relaxed clusters and often host (a fraction of them) different types of non-thermal radio emitting components, viz., 
`haloes', `mini-haloes', `relics' and `phoenix' within their Intra Cluster Medium (ICM). The radio haloes show steep ($\alpha = -1.2$) and ultra steep ($\alpha < -1.5$) spectral properties 
at low radio frequencies, giving important insights on the merger (pre or post) state of the cluster. Ultra steep spectrum radio halo emissions are rare and expected to be the 
dominating population to be discovered via LOFAR and SKA in the future. Further, the distribution of matter (morphological information), alignment of hot X-ray
emitting gas from the ICM with the total mass (dark + baryonic matter) and the bright cluster galaxy (BCG) is generally used to study the
dynamical state of the cluster.
We present here a multi wavelength study on 14 massive clusters from the CLASH survey 
and show the correlation between the state of their merger in X-ray and spectral properties (1.4 GHz - 150 MHz) at radio wavelengths. 
Using the optical data we also discuss about the gas-mass alignment, in order to understand the interplay between dark and baryonic matter in massive galaxy clusters.
\end{abstract}



\section{Introduction}
Galaxy clusters are multiple gravitationally bound systems in our Universe that exist at the dense nodal regions of the cosmic web. 
They are continuously fed with galaxy groups, sub clusters, gas, dust etc. via 
filaments, giving rise to massive systems of the order of 10$^{15}$${M_{\odot}}$. While ${\sim 70 - 80 \%}$ of the total mass in 
massive galaxy clusters is dominated by dark matter, followed by $\sim 15 - 20 \%$ of hot gas ($T \sim 10^{8} K$) in the ICM, 
the remaining is in the form of a few percent of baryonic matter in the galaxies. 
The dark matter dominates the cluster 
potential and is expected to be interacting through gravity only, thus collision less. While, the in-falling baryonic matter in the central region of galaxy 
clusters undergoes mergers that dissipate a huge amount of energy giving rise to turbulence, cold fronts and shocks within the 
ICM. The dissipated energy causes heating of the ICM gas that emits in X-rays via thermal bremsstrahlung radiation 
and accelerates the cosmic ray electrons within the ICM upto relativistic energies 
giving rise to non-thermal radio synchrotron radiation in the presence of magnetic fields (Brunetti et al. 2014). 

Depending upon the intensity of merger activity within the cluster central region, the non-thermal synchrotron emission can be classified into 
strongly polarized peripheral `relics' (a few kpc - Mpc scale) and central `Phoenix' (a few kpc scale) caused via shocks 
or compression respectively, encountered within the ICM. In addition, the turbulence caused via mergers or gas sloshing gives rise to central un polarized 
haloes ($>$ 500 kpc upto 1 Mpc scale) in unrelaxed elongated `non cool clusters' (NCC) and mini-haloes ($<$ 500 kpc) in relaxed compact `cool clusters' (CC), respectively. 
CC clusters tend to host a central bright radio loud BCG, while NCC clusters host multiple ellipticals in their center associated to each sub group under collision.
In spite of more than 100s of radio detection to-date of the diffuse synchrotron emission from the ICM in massive galaxy clusters, their origin still 
remains unclear (Ferreti et al. 2012). Thus multi wavelength study is extremely important in order to unravel the correlation between the dynamical 
state of the cluster and the origin of the non-thermal emission from their ICM. 

\section{Dynamical state and Non-thermal emission in massive galaxy clusters}
In this paper we present radio analysis on a sample of 14 massive CLASH clusters (Table 1) (Postman et al. 2012) where deep Giant Metrewave Radio Telescope (GMRT)
observations are available. The low frequency radio observations (+ archive and TGSS survey) with the GMRT down to 150 MHz has confirmed detection
of radio haloes, mini-haloes and emission from the BCG in all the 14 systems. Combined high frequency data at 1.4 GHz with the Very Large Array (VLA) has been used to derive their
spectral information. These clusters are very luminous (L$_{Bol}=10^{44}~erg~s^{-1}$, T $>$ 5 keV) at X-ray wavelength and show lensing properties at optical wavelength. 
The result of our multi wavelength analysis are listed below and in Table 1: 
\begin{table*}
\footnotesize{
\caption{Multi wavelength properties of CLASH clusters}
\label{table:3}
\centering
\begin{tabular}{c c c c c c c c c c c }     
\hline\hline
          Source   & z  &X-ray   &$L_{Bol}^{*}$    &$Temp^{*}$            &sub groups,    &Radio        &$\alpha$$_{Halo}$              \\
                   &       &morphology&$10^{44} erg s^{-1}$    &keV               &BCG            &morphology &                               \\
\hline
Non cool-core clusters\\
\hline
MACS $J0416.1-2403$&0.39  &extended  &16.0$\pm$0.9 &7.5$\pm$0.8          & 4,2         &Halo (USSR)         &$\alpha$${^{1400}_{235}}$=$-1.50\pm0.80$$^{a}$ \\
MACS $J1149.5+2223$&0.54  &extended  &30.2$\pm$1.2 &8.7$\pm$0.9          & 4,1         &Halo (Giant USSR)   &$\alpha$${^{1400}_{235}}$=$-2.10\pm0.22$$^{a}$ \\
                   &       &          & &                      &             &+ Relic           &                \\
MACS $J0717.5+3745$&0.55  &extended  &55.8$\pm$1.1 &12.5$\pm$0.7         & 5,          &Halo (Giant)         &$\alpha$${^{1400}_{235}}$=$-0.98\pm0.03$$^{a}$  \\
                   &       &          & &                      &multiple     &+ Phoenix         &                 \\
\hline
Cool-core Clusters\\ 
\hline
RX   $J1532.9+3021 $&0.36 &compact  &20.5$\pm$0.9 &5.5$\pm$0.4           & 1,1            &mini-Halo        &$\alpha$${^{1400}_{235}}$=$-1.10\pm0.07$$^{a}$\\
MACS $J0329.7-0211$&0.45  &compact  &17.0$\pm$0.6 &8.0$\pm$0.5           & 1,1            &mini-Halo (USSR) &$\alpha$${^{1400}_{610}}$=$-2.40\pm0.04$$^{a}$\\
MACS $J1931.8-2635$&0.35  &compact  &20.9$\pm$0.6 &6.7$\pm$0.4           & 1,1            &mini-Halo? (USSR)&$\alpha$${^{1400}_{150}}$=$-2.10\pm0.10$$^{b}$\\
RX   $J2129.7+0005$&0.23  &compact  &11.4$\pm$2.0 &5.8$\pm$0.4           & 1,1            &mini-Halo        &$\alpha$${^{1400}_{235}}$=$-0.60\pm0.00$$^{c}$ \\
RX   $J1347.5-1145$&0.45  &compact  &90.8$\pm$1.0 &15.5$\pm$0.6          & 1,1            &mini-Halo        &$\alpha$${^{1400}_{235}}$=$-0.92\pm0.03$$^{d}$\\
MACS $J1206.2-0847$&0.44  &compact  &43.0$\pm$1.0 &10.8$\pm$0.6          & 1,1            &mini-Halo$^{e}$? & - \\
MACS $J1115.9+0129$&0.35  &compact  &21.1$\pm$0.4 &8.0$\pm$0.4           & 1,1            &mini-Halo (USSR) &$\alpha$${^{1400}_{235}}$=$-1.80\pm0.20$$^{a}$  \\
Abell $611$        &0.29  &compact  &11.7$\pm$0.2 &7.9$\pm$0.3           & 1,1            &No detection$^{c}$& - \\
Abell $1423$       &0.21  &compact  &7.8$\pm$0.2  &7.1$\pm$0.6           & 1,1            &No detection$^{c}$& - \\
Abell $2261$       &0.22  &compact  &18.0$\pm$0.2 &7.6$\pm$0.3           & 1,1            &No detection$^{c}$& - \\
Abell $209$        &0.21  &slightly &12.7$\pm$0.3 &7.3$\pm$0.54   & 1,1            &Halo (Giant)   &$\alpha$${^{1400}_{610}}$=$-1.20\pm0.20$$^{f}$  \\
                   &       &extended &             &               &                &               & \\
\hline\hline
\end{tabular}
\footnotesize{*: Postman et al. 2012, a:Pandey-Pommier et al. 2016, 2014, 2013, b:Giacintucci et al. 2014, c:Kale et al. 2013, 2015, d:Gitti et al. 2007, 
e:Young et al. 2015, f:Venturi et al. 2008}
}
\end{table*}
\begin{itemize}
\item 3 merging NCC massive clusters hosts radio haloes that trace the extent of X-ray emission along with the shock regions 
(phoenix and relics) and radio emission from the BCGs.
\item MACS $J0416.1-2403$ shows 2 bright BCGs and multiple sub-groups under collision with an Ultra Steep Spectrum Radio (USSR) halo (Fig. 1) 
($S_{\nu}={\nu}^{\alpha}$, $\alpha<-1.5$, where $\it S$ is the flux density (mJy), ${\nu}$ the frequency (MHz) and $\alpha$ the spectral index). 
The USSRH are rare population of radio haloes that are relatively brighter at lower frequencies and caused by non-thermal 
emission from the population of old electrons after the less-common post-major merger or more frequent and less energetic pre-minor merger events in the Universe 
(Pandey-Pommier et al. 2015, Ogrean et al 2015, Cassano et al. 2010). These USSR haloes have low surface brightness and are expected to be the dominant population that will 
be discovered via new generation low frequency interferometer arrays like LOw Frequency ARray (LOFAR) and Square Kilometer Array (SKA), thanks to their high ($\mu$Jy) sensitivity 
and resolution (arcsec) at MHz-range(Cassano et al. 2015). 
\item MACS $J1149.5+2223$ shows a bright BCG in the center and multiple sub-groups undergoing collision. It hosts a possible giant USSR halo with
shocked relic region (refer Fig.1), suggesting a post violent merger phase (Bonafede et al. 2013, Pandey-Pommier et al. 2016, Golovich et al. 2016).
\item MACS $J0717.5+3745$ shows multiple bright galaxies associated to several sub-groups under collision in the center of the cluster. It hosts the most powerful 
giant radio halo ($\alpha>-1.2$) known in galaxy clusters and a shocked phoenix region at the center (Fig. 1, Pandey-Pommier et al. 2013, van Weeren et al. 2008). The 
spectral and dynamical properties of the cluster suggest that its in a state of on-going merger.
\item 11 CC massive clusters in our list either host mini-haloes or emission from the central radio loud BCG. Using high resolution observations (2") at 
1.4 GHz and low resolution 
observation down to 150 MHz, we were able to marginally differentiate the emission from the jet of the AGN as well as the mini-halo.
\item 3 CC clusters (RX$J1532.9+3021$, RX$J2129.7+0005$, RX$J1347.5-1145$) show one central bright BCG and a radio mini-halo caused due to gas sloshing 
in the central region with $\alpha>-1.2$ as shown in Fig. 2 top panel (Pandey-Pommier et al. 2016, Kale et al. 2013, 2015, Gitti et al. 2007). 
These clusters are in relaxed state with on-going minor merger activities.
\item 3 CC clusters (MACS $J1115.9+0129$, MACS$J0329.7-0211$, MACS$J1931.8-2635$)
show one central bright BCG and rare USSR mini-haloes with $\alpha<-1.5$ as seen in Fig. 2 bottom panel, indicating that these clusters are entering a post merger phase, 
with the population of old electrons mainly emitting at lower frequencies (Pandey-Pommier et al. 2016, Giacintucci et al. 2014).
\item 4 CC clusters (MACS$J1206.2-0847$, Abell $611$, Abell $1423$, Abell $2261$) show a bright BCG with no associated mini-halo emission, however 
emission from the central BCG is clearly detected in these clusters
In the case of MACS$J1206.2-0847$, diffuse radio emission was detected a few arsec away from the central BCG, suggesting the presence of a possible mini-halo, 
however deep radio observations are needed to confirm this result (Young et al. 2007, Kale et al. 2013, 2015)  
An upper limit derived from the GMRT maps are plotted in Fig. 3, top panel for the mini-haloes and the BCGs. The non detection 
of mini-haloes indicates that the cluster has entered a relaxed state with no further merger activity in process. 
\item Abell $209$ shows a giant radio halo and an associated central bright BCG, refer Fig. 3, bottom panel. This CC cluster is exceptional with 
slightly extended morphology indicating the end of a massive merger stage or a beginning of a relaxed phase (Venturi et al. 2008).
\end{itemize}

\begin{figure}[b!]
 \centering
 \includegraphics[width=1.0\textwidth,  height=0.25\textheight, clip]{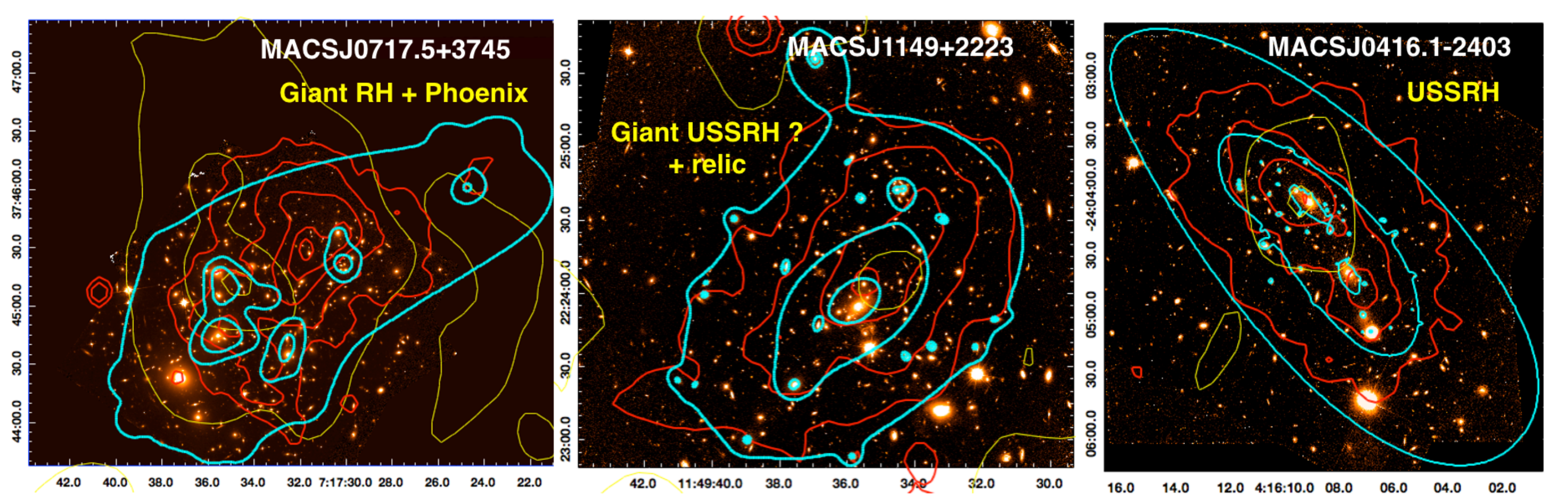}
  \caption{Dark and baryonic matter coupling in NCC clusters. X-ray (red), HST total mass map (cyan) and GMRT
(yellow) contours are overlaid on background HST map. The radio haloes are marked as `RH' and Ultra steep spectrum radio haloes as `USSRH'.
The total mass map for MACS $J0416.1-2403$ is constructed with MUSE data.}
 \label{author1:fig1}
\end{figure}

\section{Gas-mass alignment and interplay between dark and baryonic matter in massive galaxy clusters}
The multiple probes (X-ray, radio, BCG) of matter distribution in galaxy clusters can be used to study their gas-mass alignment as well as the interaction between their 
dark and baryonic matter. Dark matter dominating the
cluster gravitational potential, binds the hot, X-ray emitting and non-thermal radio emitting intra cluster gas to the cluster center. While CC clusters tend to
show homogeneous distribution in the total mass map (dark + baryonic matter, constructed from the HST/MUSE data) and positional coincidence in the 
central baryonic mass concentration versus the 
Centroid of the cluster, the NCC often show in-homogeneous total mass distribution with non-coincidence (offset) in the central baryonic mass concentration 
versus the Centroid (Limousin et al. 2012). 
This is mainly due to the violent dynamical interaction in the NCC clusters, that produces shocks or pressure waves disturbing the gas 
(Cassano et al. 2010). In the following section we will discuss the interaction of dark and baryonic matter in our sample of 14 massive CLASH clusters.

\begin{itemize}
\item The baryonic matter (X-ray + radio emitting gas and BCG) in 3 NCC massive clusters namely,
MACS $J0717.5+3745$, MACS $J1149.5+2223$ and MACS $J0416.1-2403$ shows no well defined center or positional coincidence in the multi wavelength data. 
In the case of MACS $J0416.1-2403$, the total mass map was constructed with the MUSE data providing a better constraint on the extent of the cluster mass, 
thanks to its high sensitivity to detect faint sources (refer Fig. 1, last panel). Further, a shift (a few arcsec) in the peak of optical total mass map, X-ray peak from 
the thermal gas and radio peak from the non-thermal gas in the ICM is detected, suggesting an ongoing merger activity in these clusters (refer Fig. 1). The
shift in the peak also indicates the decoupling of dark matter and baryonic matter gas components suggesting a
post-merger scenario for MACS $J1149.5+2223$ and MACS $J0416.1-2403$ clusters with USSR haloes and on-going merger scenario for MACS $J0717.5+3745$ cluster with less-
steep radio halo.
\begin{figure}[b]
 \centering
 \includegraphics[width=1.0\textwidth, height=0.25\textheight, clip]{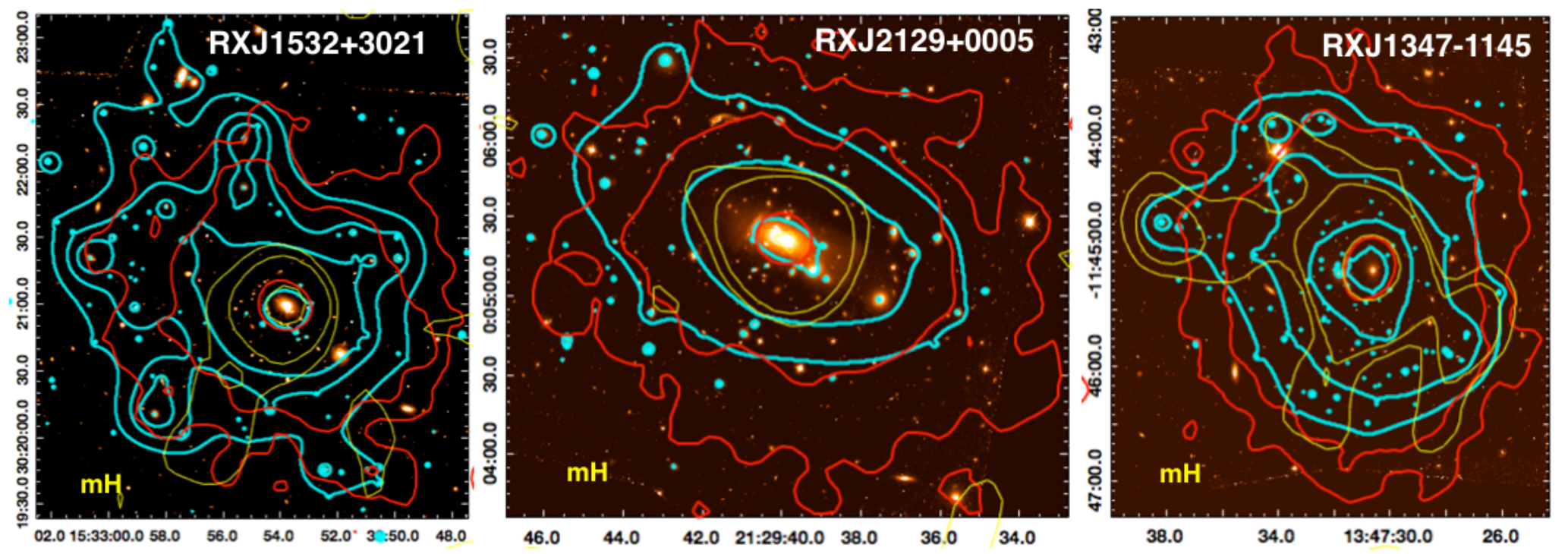}
\includegraphics[width=1.0\textwidth, height=0.25\textheight, clip]{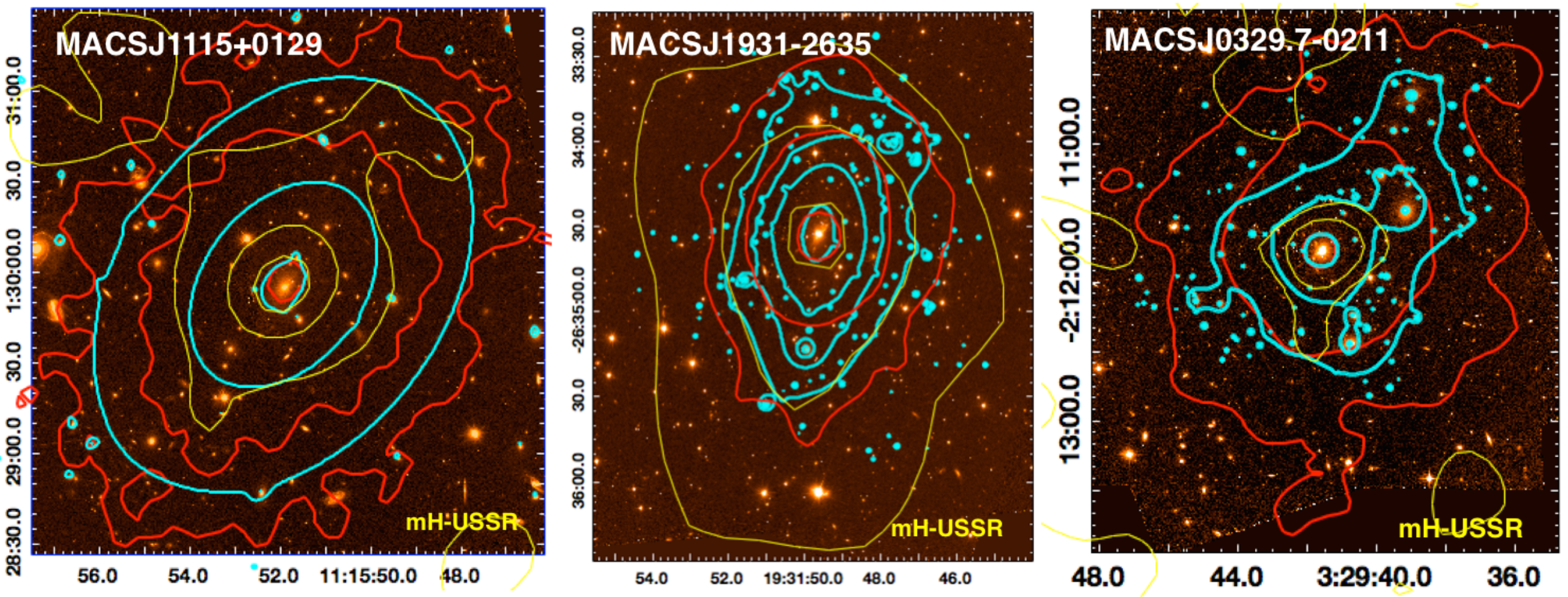}
  \caption{Dark and baryonic matter coupling in CC galaxy clusters. X-ray (red), HST total mass map (cyan) and 
GMRT (yellow) contours are overlaid on background HST map. The mini-haloes are marked as `mH' and Ultra steep spectrum radio mini-haloes as `mH-USSR'.
}
  \label{author1:fig1}
\end{figure}
\item In the case of 6 (RX $J1532+3021$, RX $J2129+0005$, RX $J1347-1145$, MACS $J1115+0129$, MACS $J1931-2635$, MACS $J0329-0211$) CC clusters the 
central region shows aligned X-ray, radio and total mass peak, suggesting that the clusters are in relaxed phase with minor merger activity and 
coupled dark and baryonic matter (Fig. 2).
\newpage
\item In the case of 4 CC clusters a coincidence in the peaks of X-ray and total mass map is seen. The radio emission from the BCG, wherever detected is aligned 
with the X-ray/total mass map peak, suggesting no more on-going merger activity (refer Fig. 3).
\begin{figure}[b]
 \centering
 \includegraphics[width=1\textwidth, height=0.5\textheight, clip]{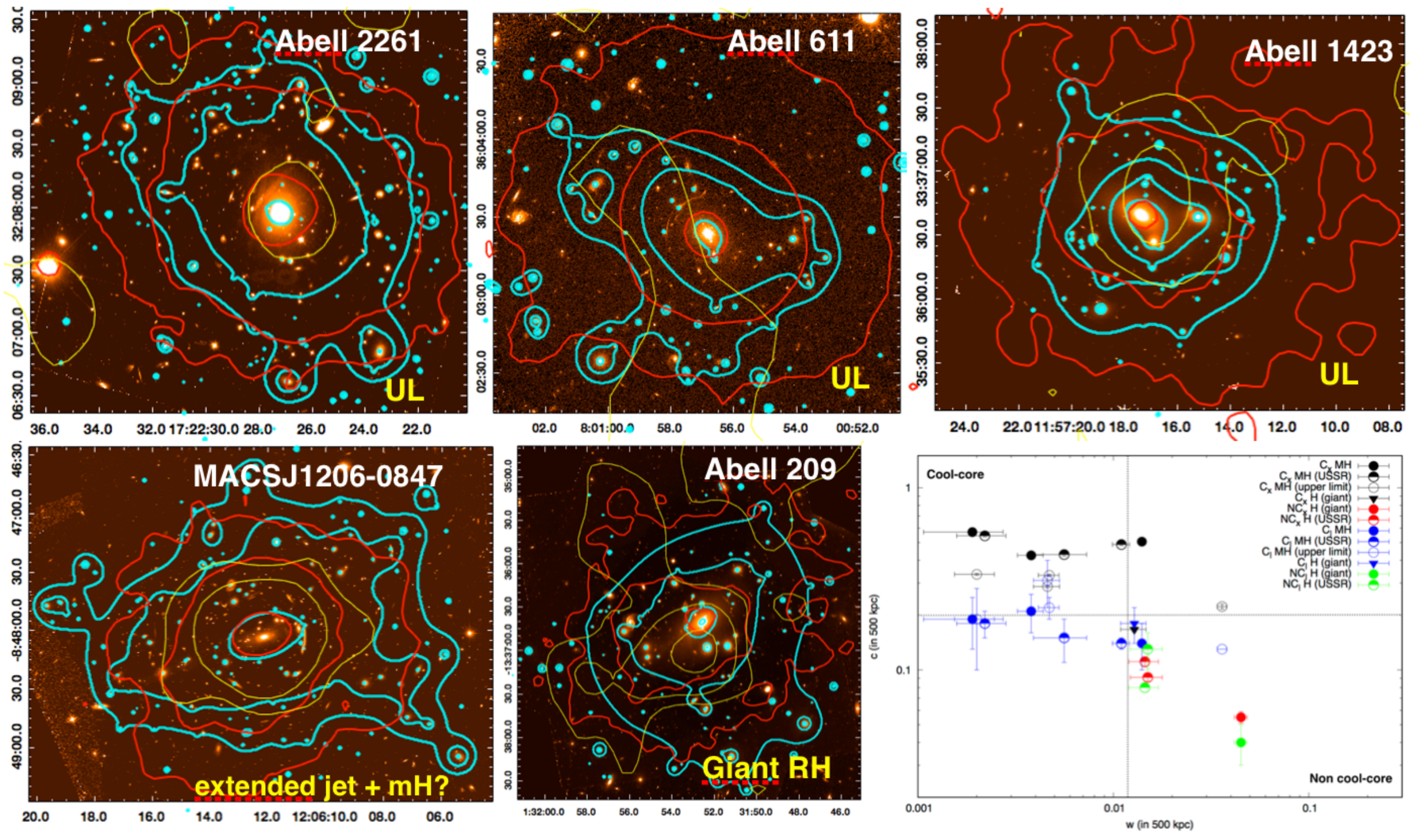}
  \caption{Dark and baryonic matter coupling in CC galaxy clusters. X-ray (red), HST total mass map (cyan) and
GMRT (yellow) contours are overlaid on background HST map. The non-detection upper limits are marked as `UL', radio halo as `RH', X-ray concentration in 
cool-core as `C$_x$', X-ray concentration in non cool-core as `NC$_x$', Concentration in total mass map via lensing analysis as `C$_l$' in cool-cores and `NC$_l$' 
in non cool-core clusters.}
  \label{author1:fig1}
\end{figure}
\item Finally, the exceptional CC cluster Abell $209$ hosting a giant radio halo shows an offset in 
the peak of radio and total mass map with a coincidence in the peak of total mass and X-ray (Fig. 3, bottom middle panel), suggesting its transition stage from the 
NCC to CC phase. The cluster appears to be cooling down after a violent merger phase.
\item The morphological and spectral information from the multi wavelength data was used to infer the state of relaxation of galaxy clusters as well as the interplay 
between their dark and baryonic matter. The X-ray and total mass map concentration (ratio between the light within a circular aperture of minimum 100 kpc 
and maximum 500 kpc) of each cluster in our list is plotted with respect to the shift in their X-ray 
Centroid (standard deviation of the projected separation between the X-ray peak and Centroid estimated within circular aperture of maximum 500 kpc) in Fig. 3, 
bottom right panel. It is clear from the figure that the concentration of CC clusters is higher near the central region and they show smaller 
shift in their X-ray Centroid compared to the NCC clusters, confirming their non relaxed nature. Further, the mass concentration in X-rays is higher in general for 
both the CC and NCC clusters compared to the total mass map (with dominating dark matter) derived from the optical data, confirming the collisional nature of the 
baryonic matter traced via X-ray observations at the cluster center. The lower mass concentration at the cluster center in the total mass map suggests its 
collision less nature, thus minimally affected via merger events.     
\end{itemize}

\section{Conclusions}
Massive galaxy clusters provide the most interesting astrophysical laboratories to study the merger activity and emission processes in the dense regions of
the Universe. In this paper, we present a multi wavelength (GMRT, Chandra, HST/MUSE) analysis of morphological and spectral properties on a sample of 14 massive galaxy clusters in 
the CLASH catalog. The multi wavelength data was used to derive a correlation between the dynamical state and the non-thermal emission from the cluster ICM and study the dark 
and baryonic matter interaction in their central region. The main results derived in this paper are listed below:
\begin{itemize}
\item Low frequency observations (down to 235 or 150 MHz) not only confirm the presence of mini-haloes and haloes in massive galaxy clusters but also provide information 
about their dynamical state.
\item Radio morphology gives the information about the
violent or less energetic merger activity in the cluster and the spectral index gives information about the pre-
or post- merger state of the cluster.
\item The radio (non-thermal) halo and X-ray (thermal) emission are generally coincident and originate from the ICM.
\item The peak in mass distribution at X-ray and total mass map is coincident for CC clusters while non-coincident in NCC clusters.
\item An offset in the peak of radio and total mass map as well as X-ray is detected in clusters hosting haloes, where
the merger is still active. 
\item Combined multi wavelength analysis of the mass distribution in galaxy clusters suggests that dark matter is loosely interacting 
while the baryonic matter is tightly interacting during merger events in the cluster central regions. 
\end{itemize}

\begin{acknowledgements}
We thank the staff of GMRT, who made these observations possible. The GMRT is run by the National Centre for Radio Astrophysics of the Tata Institute of Fundamental 
Research. The authors are thankful to the research grant received from the Franco-Indian CEFIPRA organization. This research has made extensive use of the data 
available from the public archive from HST, Chandra, VLA and GMRT observatory. MP thanks Dr. Adi Zitrin for useful discussion.  
\end{acknowledgements}


\bibliographystyle{aa}  
\bibliography{sf2a-template} 
\end{document}